\documentstyle[12pt]{article}

\begin{document}

\begin{titlepage}
\begin{center}
\large\bf Freeze-out configuration in multifragmentation \\
\normalsize\bf
\vspace{1.0cm}
Bao-An Li$^{a}$, D.H.E. Gross$^{a*}$, V. Lips$^{b}$ and H. Oeschler$^{b}$\\
\vspace{0.5cm}
\normalsize
a. Hahn-Meitner-Institut, Bereich Schwerionenphysik\\
 Glienickerstr.100, D-14109 Berlin, Germany\\
 $^{*}$ and Fachbereich Physik der Freien Universit\"at,
 Berlin, Germany \\
\vspace{0.3cm}
b. Institut f\"ur Kernphysik\\
Technische Hochschule Darmstadt, D-64289 Darmstadt, Germany
\end{center}
\date
\vspace{0.5cm}

\begin{quote}
The excitation energy and the nuclear density at the time of breakup
are extracted for the $\alpha + ^{197}Au$ reaction at beam energies of
1 and 3.6 GeV/nucleon. These quantities are calculated from
the average relative velocity of intermediate
mass fragments (IMF) at large correlation angles
as a function of the multiplicity of IMFs using a statistical
model coupled with many-body Coulomb trajectory calculations.
The Coulomb component $\vec{v}_{c}$ and thermal component
$\vec{v}_{0}$ are found to depend oppositely on the excitation energy,
IMFs multiplicity, and freeze-out density.
These dependencies allow the determination of both
the volume and the mean excitation energy at the time of breakup.
It is found that the volume remained constant as the beam
energy was increased, with a breakup density of about $\rho_{0}/7$,
but that the excitation energy increased $25\%$ to about 5.5 MeV/nucleon.
\end{quote}

\begin{quote}
PACS Number 25.70.Np
\end{quote}

\vspace{1.0cm}


\end{titlepage}

\begin{doublespacing}

In spite of intensive experimental and theoretical efforts,
the mechanism of nuclear multifragmentation
remains a subject of much interest and debate.
The study becomes even more attractive
with the availability of more high quality data from powerful $4\pi$
detectors \cite{moretto}.

In this Letter we report on results of the
determination of the freeze-out density and excitation energy by
studying the average relative velocity \\
$<v_{rel}>$ of IMFs at large correlation angles as a function of
the multiplicity of IMFs. We use the Berlin
version of a statistical model \cite{gross90} coupled with the
many-body Coulomb trajectory calculations to describe the
multifragmentation process. Among the many available multifragmentation
models, those based on the simultaneous breakup mechanism have been
rather successful in explaining a number of aspects of multifragmentation
data \cite{gross92}. Moreover, statistical phase-space models like ours
have a unique ability for calculating many-fragment
correlation functions with reliable statistics.

In the simultaneous breakup scenario, one can picture the
multifragmentation as follows: A composite nuclear system
formed initially in any reaction will expand under thermal
and/or compressional pressure. The expansion may take the
system into the hydrodynamical instability region where the
sound velocity is imaginary\cite{bertsch}. In the
instability region the system breaks up into fragments and
nucleons. At the same time the further expansion slows down due to the
creation of new surfaces of fragments and becomes chaotic due to
the strong interactions among fragments and nucleons through the
exchange of nucleons and momenta.  At the end of this chaotic
expansion (the instant of freeze-out) the fragments have
rearranged themselves such that the entropy of the system has reached
its maximum. At the moment of the freeze out, the momentum
distribution, spatial distribution, mass distribution and
internal excitations of fragments of the system are solely
determined microcanonically by the available phase space. After
the freeze out, fragments and nucleons move along their Coulomb
trajectories. This picture is in agreement with a recent BUU-model
simulation of the reaction dynamics leading to
multifragmentation\cite{grossli,li}.

Within the above picture, the most important quantities
determining the final state of the multifragmentation events are the
excitation energy and the density or size of the system at the
freeze out. Ideally, these quantities can be calculated from
dynamical models for nuclear reactions. However, the breakup
conditions (time scale, entropy and partitioning into fragments) is
the most interesting topic of current research because none of
the existing dynamical models can handle the breakup properly.
The extracted freeze-out density and excitation energy will
put strong constraints on the reaction dynamics leading to the
formation of hot nuclei and on the nuclear equation of state for
low-density nuclear matter. Moreover, to a significant extent,
our study itself is a study of the breakup dynamics.

The velocities of the fragments in the final state are sensitive
to their spatial distribution at the freeze out due to their
long range Coulomb repulsion. The velocities are also sensitive to the
excitation energy of the fragmenting system. The charge and
mass distributions of fragments, as well as the thermal velocities
of the fragments at the freeze out, are mainly determined by the
excitation energy of the system. Therefore, the
relative velocity distributions are useful tools for studying the
breakup process as well as for extracting the freeze-out density
and excitation energy. The above discussions show that the two
quantities have to be extracted simultaneously. It
is interesting to mention that the relative velocity
distributions for various fragment classes have been used to
distinguish between the simultaneous breakup scenario and the
sequential binary decay scenario
\cite{tro87,gross89,poch89,barz90} with contradictory
conclusions. By selecting dedicated event classes a deeper
insight can be gained. For three-body events from the system Ar +
Au at 30 MeV/nucleon it was shown that the relative velocity
distribution can be well reproduced assuming the sequential
binary decay of a hot system at an excitation energy of about
3~MeV/nucleon\cite{bizard92}. This result can be understood as a
deep inelastic collision, followed by the fission of the heavy
partner \cite{lip93}. At higher incident energies, selecting
events involving only three heavy fragments, the ALADIN
collaboration has very recently also studied the relative
velocity distribution from 600 MeV/nucleon Au induced reactions
\cite{linden93}, their results suggest a fast disintegration of
the entire highly excited system. A range of the freeze-out radii
and excitation energies was extracted simultaneously using a
simple fragmentation model coupled with three-body trajectory
calculations.

As a first part of this Letter we study the model predictions for
the average relative velocity between IMFs with large correlation
angles as a function of the total IMF multiplicity, varying the
freeze-out density and the excitation energy. In the second part we
compare these results with experimental data in order to deduce
the freeze-out condition. This method goes far beyond the usual
way of comparing the relative velocity distribution for events
involving only three fragments or the relative velocity
distribution obtained without any constraint on the multiplicity
of IMFs. It is evident that the multiplicity and charge
of the unobserved fragments have decisive influence on the
observed velocity distributions. We want to emphasize that this
means we test more than just a three-body dynamics as in ref.\cite{linden93}.
The change of $<v_{rel}>$ with the IMF multiplicity
at a relatively constant excitation energy gives the clue to our method
of extracting the source size and excitation energy. This is a clear
advantage using the very asymmetric collision system $\alpha + Au$.
There is also a qualitative difference between
highly energetic $\alpha$ induced fragmentations and heavy-ion induced ones.
In the later the variation of the IMF multiplicity is mainly due to the broad
variation of the excitation energy. Here it is much more due to
the {\em fluctuation} of $M_{IMF}$ at a relatively narrow distribution of
excitation energies.

The final velocity of each fragment $\vec{v_{f}}$
can be decomposed as \[ \vec{v_{f}}=\vec{v_{0}}+\vec{v_{c}}, \]
where the $\vec{v_{0}}$ term is the thermal velocity of the
fragment at the freeze out. At a fixed excitation energy of the
fragmenting system the $\vec{v}_{0}$ term increases with the
multiplicity of IMFs due to the smooth decrease of the average mass of
the IMFs. The $\vec{v_{c}}$ term is the velocity of the fragment
gained after moving in the Coulomb field of the rest of the
system and thus depends on the radius of the system at the freeze out.
The dependence of $\vec{v}_{c}$ on the multiplicity of IMFs is
somewhat more complicated. It is strongly influenced by the existence
of a third, larger fragment. This will be discussed in detail later.
Therefore, the dependence of $<v_{rel}>$ on the
multiplicity of IMFs results from the interplay between the two
velocity components of the IMFs just mentioned. They reflect the
freeze-out density and the excitation energy. The secondary
deexcitation of primary hot fragments will slightly alter the
relative velocity distribution. Since the deexcitation occurs
late, the influence of the change in the Coulomb field is small.
Further, since the emission is assumed
to be isotropic, only the width will be changed, but
not the mean value.

Necessary inputs to our calculations are the mass and the charge
of the fragmenting source. Since this Letter relates to $\alpha$~+~Au
reactions at 1 and 3.6~GeV/nucleon, cascade model
calculations \cite{toneeve} were performed for these reactions,
yielding the mass and charge of the residual nucleus, which is
the emitting equilibrized source. For the higher incident energy a
mean residue mass of A=150 and charge of Z=62 was
deduced, for the lower incident energy the mean residue mass and
charge was determined to be A=170 and Z=70, respectively.  The mass
and charge of the heavy residue are rather independent of the
impact parameter due to the large mass asymmetry. This
legitimates the comparison of the experimental
results to the statistical calculations using a fixed residue
mass, charge and excitation energy. This is an important
advantage of the highly asymmetric $\alpha$~+~Au system for
investigations of this kind.

To see how the dependence of $<v_{rel}>$ on the multiplicity of
IMFs is sensitive to the breakup conditions of hot nuclei towards
multifragmentation, we study in Fig.\ 1~(a) the
excitation energy dependence of $<v_{rel}>$ at large correlation
angles for a given breakup density.

In Fig.\ 1 (a)-(c), the calculations are done for excitation
energies of 4, 5, 8 and 10 MeV/nucleon respectively at a fixed
freeze-out radius of $2.1A^{1/3}$ and a residue mass and charge
of A=150 and Z=62. It is seen that for small excitation energies
$<v_{rel}>$ decreases with increasing multiplicity of the intermediate
mass fragments $M_{IMF}$ and that $<v_{rel}>$
is higher at $E^{*}$~=4 MeV/nucleon than at 5 MeV/nucleon.  However, an
opposite behaviour is seen at high excitation energies. Now
$<v_{rel}>$ increases with increasing $M_{IMF}$ and $<v_{rel}>$
is higher at 10 MeV/nucleon than at 8~MeV/nucleon. These features
of $<v_{rel}>$ can be understood by considering the interplay
between the two velocity components of the fragments and their
excitation energy dependence. For this purpose we show in Fig.~1~(b)
the average maximum charge of the fragments as a function of the
multiplicity of IMFs at the corresponding excitation energies. The
dashed line at Z=15 is the upper boundary of the charge defined
for the IMFs. At lower excitation energies (e.g. 4 and 5 MeV/nucleon) the
average maximum charge is larger than 15 for a significant range
(if not all) of the IMF multiplicities.  The fragment with the
highest charge in each event accelerates the other fragments and
is therefore responsible for the higher values of $<v_{rel}>$.
The maximum charge of these events decreases strongly with
increasing IMF multiplicity. This explains the trend of
decreasing $<v_{rel}>$ with increasing $M_{IMF}$ observed at
lower excitation energies.

At high excitation energies the maximum charge is within the
charge range of IMFs for all statistically significant events
(seen from Fig.~1~(c)) and the thermal term $\vec{v}_{0}$
dominates over the Coulomb term $\vec{v}_{c}$.  Since the average
mass of IMFs will decrease as the multiplicity of IMFs increases,
the $\vec{v}_{0}$ term and
consequently $<v_{rel}>$ of IMFs will increase with increasing
$M_{IMF}$.  The fact that $<v_{rel}>$ is higher at higher
excitation energies is due to both the higher temperature reached
and the lower average mass of the fragments. To evaluate the
statistical significance of each multiplicity of IMFs, we have
presented the distribution of the multiplicity of IMFs at the
corresponding excitation energies in Fig.~1~(c). The typical
rise and fall behaviour of the IMF multiplicities as a function of
the excitation energy can be seen between $E^{*}$ = 5 and 10
MeV/nucleon. The average IMF multiplicity at $E^{*}$ = 10
MeV/nucleon is lower than that at 8 MeV/nucleon. This is due to
the increasing dissociation into light fragments (Z $<$ 3)  and
nucleons at higher excitation energies.

To see the effects of the freeze-out density or radius on $<v_{rel}>$ we
show in Fig.\ 1 (d) $<v_{rel}>$ at large correlation angles as a
function of $M_{IMF}$ for freeze-out radii of $2.1A^{1/3},
2.4A^{1/3}$ and $2.6A^{1/3}$ respectively at a fixed excitation
energy of 5 MeV/nucleon. The magnitude of $<v_{rel}>$ increases
rapidly with decreasing freeze-out radius due to the stronger
Coulomb repulsion at smaller freeze-out radii. It is interesting
to note that at $R_{c}=2.6A^{1/3}$ the Coulomb repulsion is so
weak that for $M_{IMF}\geq 6$ the thermal term $\vec{v}_{0}$
starts dominating over the Coulomb term $\vec{v}_{c}$ and $<v_{rel}>$
starts increasing with increasing $M_{IMF}$.

After establishing the various dependencies of $<v_{rel}>$ we can
compare them with the experimental data. For this purpose the reaction
$\alpha$~+~Au is ideally suited, since all fragments originate
from one source, the target residue. Additionally the recoil
velocities are small which facilates the determination of the
velocity and angular correlations with high precision.
The reaction has been studied at 1 and 3.6~GeV/nucleon incident
energy at the synchrophasotron of the JINR in Dubna using the new
$4\pi$ setup ``FASA''\cite{avd93}. In \cite{lip93a} it has been demonstrated
that sufficient excitation energy is reached to
induce the multifragment breakup of the system into many IMFs.

The multiplicity of the IMFs was measured by a fragment multiplicity
detector (FMD) system covering a large part of the $4\pi$ solid
angle. The energy, velocity and mass of single fragments from the
event were determined with high precision using time-of-flight
telescopes (TOF). The relative angle and velocity of coincident
fragments were measured using a position-sensitive parallel-plate
avalanche counter (PPAC). For the FMD and the PPAC an IMF is
defined as a fragment with a charge number of $3 \leq Z \leq 15$.
For TOF it is defined correspondingly
as a fragment with a mass number of $6~\leq~A~\leq~30$. The large
correlation-angle data were taken by three TOF-PPAC combinations
covering the correlation angular range of
$105^{\circ}-155^{\circ}, 130^{\circ}-180^{\circ}$ and
$150^{\circ}-180^{\circ}$ respectively. Efficiencies of both the
TOFs and PPACs are taken into account in the data analysis and
the model calculations.

The comparison to the experimental data
is shown in Fig.\ 2, which gives the average relative velocity of the
IMFs at large relative angles, detected in TOF and PPAC, as a
function of the multiplicity of the IMFs measured in the FMD.
In this figure the efficiency of the FMD
was not corrected but considered in the respective calculations.
Figure 2 shows, that a combination of a freeze-out density
of $\rho_{0}/7$ and an excitation energy of 4.5 MeV/nucleon for
the lower incident energy and the same freeze-out density but a
higher excitation energy of 5.5 MeV/nucleon for the higher incident
energy can reproduce the experimental data reasonably well. With this set
of parameters, our calculations also well reproduce many other
aspects of the experiment simultaneously, including the average
IMF multiplicity, the width and the shape of the relative
velocity distribution as well as the mass distribution of
fragments. To estimate fluctuations of $<v_{rel}>$
originating from fluctuations of the residue charge and excitation energy
we have performed calculations for the case of $E_{beam}/A=3.6$ GeV
with Z=62, 70 and 79 respectively. By adjusting the excitation energy
between 5.5 and 6.6 MeV/nucleon in order to have a similar average
IMF multiplicitiy of about 5.9, the $<v_{rel}>$ varies by about $4\%$
which corresponds to a variation of $R_{c}$ by about $5\%$,
yielding an uncertainty of the breakup density of $15\%$.

In summary, using a statistical model coupled with many-body
Coulomb trajectory calculations we have studied the average
relative velocity of intermediate mass fragments as a function of
the multiplicity of IMFs at different but constant excitation energies and
freeze-out radii of the fragmenting source. These two quantities can be
determined individually because of the complex
and different forms of the dependence of $<v_{rel}>$ on the IMF
multiplicity at different excitation energies and freeze-out
densities. For the $\alpha + ^{197}$Au system at beam energies of 1 and 3.6
GeV/nucleon, the freeze-out density was found to be
$\rho_{c}\approx\rho_{0}/7$.
While the mean excitation energy increases from 4.5 MeV/nucleon to
5.5 MeV/nucleon. The freeze-out density found is very
similar to the ones deduced from nucleus-nucleus collisions.
Conceivable compression effects in the later seem therefore to
play a minor role if at all.

\end{doublespacing}

\section*{Figure Captions}

\begin{description}

  \item{\bf Fig.\ 1}\ \ \ (a) Average relative velocity of
IMFs at large correlation angles as a function of the
multiplicity of IMFs, (b) the average maximum charge
of fragments as a function of the multiplicity of IMFs, (c) the
multiplicity distribution of IMFs, and (d) average relative velocity
as function of the multiplicity of IMFs for different freeze-out
radii.
  \item{\bf Fig.\ 2}\ \ \ Dependence of the average relative
velocity of IMFs at large correlation angles on the multiplicity of
IMFs measured in the FMD (not corrected for efficiency)
for the system $\alpha$~+~Au at 1.0 GeV/nucleon (filled squares: experiment,
solid line: calculation with $E^{*}/A=4.5 MeV$ and $R_{c}=2.3*A^{1/3}$)
and 3.6 GeV/nucleon (filled circles: experiment, dashed line:
calculation with $E^{*}/A=5.5 MeV$ and $R_{c}=2.3*A^{1/3}$) respectively.

\end{description}

\end{document}